\documentstyle[twoside]{article}

\catcode`\@=11
\long\def\@makefntext#1{
\protect\noindent \hbox to 3.2pt {\hskip-.9pt
$^{{\eightrm\@thefnmark}}$\hfil}#1\hfill}		

\def\thefootnote{\fnsymbol{footnote}}
\def\@makefnmark{\hbox to 0pt{$^{\@thefnmark}$\hss}}	

\def\ps@myheadings{\let\@mkboth\@gobbletwo
\def\@oddhead{\hbox{}
\rightmark\hfil\eightrm\thepage}
\def\@oddfoot{}\def\@evenhead{\eightrm\thepage\hfil
\leftmark\hbox{}}\def\@evenfoot{}
\def\sectionmark##1{}\def\subsectionmark##1{}}

\oddsidemargin=\evensidemargin
\renewcommand{\thefootnote}{\fnsymbol{footnote}}

\newcounter{sectionc}
\newcounter{subsectionc}
\newcounter{subsubsectionc}
\renewcommand{\section}[1] {\vspace{12pt}\addtocounter{sectionc}{1}
\setcounter{subsectionc}{0}\setcounter{subsubsectionc}{0}\noindent
	{\tenbf\thesectionc. #1}\par\vspace{5pt}}
\renewcommand{\subsection}[1] {\vspace{12pt}
\addtocounter{subsectionc}{1}\setcounter{subsubsectionc}{0}\noindent
	{\bf\thesectionc.\thesubsectionc.
        {\kern1pt \bfit #1}}\par\vspace{5pt}}
\renewcommand{\subsubsection}[1] {\vspace{12pt}
\addtocounter{subsubsectionc}{1}\noindent
        {\tenrm\thesectionc.\thesubsectionc.\thesubsubsectionc.
	{\kern1pt \tenit #1}}\par\vspace{5pt}}

\newcounter{appendixc}
\newcounter{subappendixc}[appendixc]
\newcounter{subsubappendixc}[subappendixc]
\renewcommand{\thesubappendixc}{\Alph{appendixc}.
        \arabic{subappendixc}}
\renewcommand{\thesubsubappendixc}{\Alph{appendixc}.
        \arabic{subappendixc}.\arabic{subsubappendixc}}

\renewcommand{\appendix}[1] {\vspace{12pt}
        \refstepcounter{appendixc}
        \setcounter{figure}{0}
        \setcounter{table}{0}
        \setcounter{lemma}{0}
        \setcounter{theorem}{0}
        \setcounter{corollary}{0}
        \setcounter{definition}{0}
        \setcounter{equation}{0}
        \renewcommand{\thefigure}{\Alph{appendixc}.\arabic{figure}}
        \renewcommand{\thetable}{\Alph{appendixc}.\arabic{table}}
        \renewcommand{\theappendixc}{\Alph{appendixc}}
        \renewcommand{\thelemma}{\Alph{appendixc}.\arabic{lemma}}
        \renewcommand{\thetheorem}{\Alph{appendixc}.\arabic{theorem}}
        \renewcommand{\thedefinition}{\Alph{appendixc}.
         \arabic{definition}}
        \renewcommand{\thecorollary}{\Alph{appendixc}.
         \arabic{corollary}}
        \renewcommand{\theequation}{\Alph{appendixc}.
         \arabic{equation}}
        \noindent{\tenbf Appendix \theappendixc #1}\par\vspace{5pt}}
\newcommand{\subappendix}[1] {\vspace{12pt}
        \refstepcounter{subappendixc}
        \noindent{\bf Appendix \thesubappendixc. {\kern1pt \bfit #1}}
	\par\vspace{5pt}}
\newcommand{\subsubappendix}[1] {\vspace{12pt}
        \refstepcounter{subsubappendixc}
        \noindent{\rm Appendix \thesubsubappendixc.
        {\kern1pt \tenit #1}}\par\vspace{5pt}}

\newcommand{\textlineskip}{\baselineskip=13pt}
\newcommand{\smalllineskip}{\baselineskip=10pt}

\def\eightcirc{
\begin{picture}(0,0)
\put(4.4,1.8){\circle{6.5}}
\end{picture}}
\def\eightcopyright{\eightcirc\kern2.7pt\hbox{\eightrm c}}

\newcommand{\pub}[1]{{\begin{center}\footnotesize\smalllineskip
	Preprint No. #1\\
	\end{center}
	}}

\def\abstracts#1#2#3{{
	\centering{\begin{minipage}{4.5in}\baselineskip=10pt
        \footnotesize
	\parindent=0pt #1\par
	\parindent=15pt #2\par
	\parindent=15pt #3
	\end{minipage}}\par}}

\renewenvironment{thebibliography}[1]
	{\frenchspacing
	 \ninerm\baselineskip=11pt
	 \begin{list}{\arabic{enumi}.}
	{\usecounter{enumi}\setlength{\parsep}{0pt}
	 \setlength{\leftmargin 12.7pt}{\rightmargin 0pt}
	 \setlength{\itemsep}{0pt} \settowidth
	{\labelwidth}{#1.}\sloppy}}{\end{list}}

\newcounter{itemlistc}
\newcounter{romanlistc}
\newcounter{alphlistc}
\newcounter{arabiclistc}

\newcommand{\fcaption}[1]{
        \refstepcounter{figure}
        \setbox\@tempboxa = \hbox{\footnotesize Fig.~\thefigure. #1}
        \ifdim \wd\@tempboxa > 5in
           {\begin{center}
        \parbox{5in}{\footnotesize\smalllineskip Fig.~\thefigure. #1}
            \end{center}}
        \else
             {\begin{center}
             {\footnotesize Fig.~\thefigure. #1}
              \end{center}}
        \fi}

\newcommand{\tcaption}[1]{
        \refstepcounter{table}
        \setbox\@tempboxa = \hbox{\footnotesize Table~\thetable. #1}
        \ifdim \wd\@tempboxa > 5in
           {\begin{center}
        \parbox{5in}{\footnotesize\smalllineskip Table~\thetable. #1}
            \end{center}}
        \else
             {\begin{center}
             {\footnotesize Table~\thetable. #1}
              \end{center}}
        \fi}

\def\@citex[#1]#2{\if@filesw\immediate\write\@auxout
	{\string\citation{#2}}\fi
\def\@citea{}\@cite{\@for\@citeb:=#2\do
	{\@citea\def\@citea{,}\@ifundefined
	{b@\@citeb}{{\bf ?}\@warning
	{Citation `\@citeb' on page \thepage \space undefined}}
	{\csname b@\@citeb\endcsname}}}{#1}}

\newif\if@cghi
\def\cite{\@cghitrue\@ifnextchar [{\@tempswatrue
	\@citex}{\@tempswafalse\@citex[]}}
\def\citelow{\@cghifalse\@ifnextchar [{\@tempswatrue
	\@citex}{\@tempswafalse\@citex[]}}
\def\@cite#1#2{{$\null^{#1}$\if@tempswa\typeout
	{IJCGA warning: optional citation argument
	ignored: `#2'} \fi}}

\def\pmb#1{\setbox0=\hbox{#1}
	\kern-.025em\copy0\kern-\wd0
	\kern.05em\copy0\kern-\wd0
	\kern-.025em\raise.0433em\box0}

\def\fnt#1#2{\footnotetext{\kern-.3em
	{$^{\mbox{\scriptsize #1}}$}{#2}}}

\def\fpage#1{\begingroup
\voffset=.3in
\thispagestyle{empty}\begin{table}[b]\centerline{\footnotesize #1}
	\end{table}\endgroup}

\font\tenrm=cmr10
\font\tenit=cmti10
\font\tenbf=cmbx10
\font\bfit=cmbxti10 at 10pt
\font\ninerm=cmr9

\font\eightrm=cmr8

\def\qed{\hbox{${\vcenter{\vbox{		
   \hrule height 0.4pt\hbox{\vrule width 0.4pt height 6pt
   \kern5pt\vrule width 0.4pt}\hrule height 0.4pt}}}$}}

\renewcommand{\thefootnote}{\fnsymbol{footnote}}

\input epsf
\global\arraycolsep=2pt
\def\spose#1{\hbox to 0pt{#1\hss}}
\def\lsim{\mathrel{\spose{\lower 3pt\hbox{$\mathchar"218$}}
 \raise 2.0pt\hbox{$\mathchar"13C$}}}
\def\gsim{\mathrel{\spose{\lower 3pt\hbox{$\mathchar"218$}}
 \raise 2.0pt\hbox{$\mathchar"13E$}}}

\renewcommand{\theequation}{\thesection.\arabic{equation}}
\def\laq{\raise 0.4ex\hbox{$<$}\kern -0.8em\lower 0.62
ex\hbox{$\sim$}}
\def\gaq{\raise 0.4ex\hbox{$>$}\kern -0.7em\lower 0.62
ex\hbox{$\sim$}}

\def\beq{\begin{equation}}
\def\eeq{\end{equation}}
\def\bea{\begin{eqnarray}}
\def\eea{\end{eqnarray}}

\def \pa {\partial}
\def \ra {\rightarrow}

\def \fb {\overline \phi}

\def \ti {\tilde}
\def \la {\lambda}
\def \ls {\lambda_s}
\def \La {\Lambda}

\def \b {\beta}
\def \a {\alpha}
\def \ap {\alpha^{\prime}}

\def \sg {\sigma}
\def \da {\delta}
\def \ep {\epsilon}

\def \Om {\Omega}
\def \noi {\noindent}

\begin{document}

\begin{titlepage}

\begin{flushright}
DFTT-35/97\\
gr-qc/9706037
\end{flushright}

\vspace{3 cm}

\begin{center}
\Large\bf Birth of the Universe in String Cosmology
\end{center}

\vspace{2cm}

\begin{center}
M. Gasperini\\
{\sl Dipartimento di Fisica Teorica, Universit\`a di Torino,}\\
{\sl Via P. Giuria 1, 10125 Turin, Italy}\\
and\\
{\sl Istituto Nazionale di Fisica Nucleare, Sezione di Torino, Turin, Italy}\\
\end{center}

\vspace{2cm}

\begin{abstract}
\noi
The decay of the string perturbative vacuum into our present
cosmological state is associated to the transition from a phase of
growing curvature and growing dilaton, to a phase of decreasing 
curvature and frozen dilaton. The possible approaches to a classical and
quantum description of such a transition are introduced and briefly
discussed.

\end{abstract}

\vspace{2cm}
\begin{center}
------------------------------

\vspace{2cm}
To appear in \\
{\sl Proc. of the Euroconference ``Fourth Paris Cosmology Colloquium"}\\
Observatoire de Paris, June 1997 -- 
ed. by H. De Vega and N. Sanchez \\
(World Scientific, Singapore)
\end{center}
 \vspace{1.5cm}
\vfill

\end{titlepage}


\normalsize\textlineskip
\thispagestyle{empty}
\setcounter{page}{1}


\vspace*{0.11truein}

\fpage{1}

\centerline{\bf BIRTH OF THE UNIVERSE IN STRING COSMOLOGY}
\vspace*{0.27truein}

\centerline{\footnotesize MAURIZIO GASPERINI}
\vspace*{0.015truein}
\centerline{\footnotesize\it Dipartimento di Fisica Teorica, 
Universit\`a di Torino,}
\baselineskip=10pt
\centerline{\footnotesize  {\it Via P. Giuria 1, 10125, Turin, Italy}}
\baselineskip=10pt
\centerline{\footnotesize and {\it Istituto Nazionale di Fisica Nucleare,
Sezione di Torino, Turin, Italy}}

\vspace*{0.3truein}
\abstracts
{The decay of the string perturbative vacuum into our present
cosmological state is associated to the transition from a phase of
growing curvature and growing dilaton, to a phase of decreasing 
curvature and frozen dilaton. The possible approaches to a classical and
quantum description of such a transition are introduced and briefly
discussed.  }
{}{}
\vspace*{0.225truein}
\pub{DFTT-35/97;~~~~~~ E-print Archives: gr-qc/9706037}
\vspace*{0.8pt}\textlineskip

\textheight=7.8truein
\setcounter{footnote}{0}
\renewcommand{\thefootnote}{\alph{footnote}}

\vspace*{0.125truein}

\renewcommand{\theequation}{1.\arabic{equation}}
\setcounter{equation}{0}
\section{Introduction}
\label{sec:1}
\noindent
In the context of the pre-big bang cosmological scenario\cite{1},
based on the string effective action, the process of ``birth of the
Universe" corresponds to the transition from the string perturbative
vacuum to the standard radiation-dominated regime, passing through a
high-curvature and strong coupling phase, in which quantum and
``stringy" effects may become important. Such a process may be
qualitatively illustrated as in Fig. 1, 
by plotting the time evolution of the
curvature scale from the initial vacuum down to the present
cosmological state. The aim of this paper is to introduce, and briefly
discuss, the classical and quantum approach to the transition from the
pre-big bang to the post-big bang regime. 

Before starting the discussion, let me stress that the physical system
that we call Universe evolves in time, like all 
physical systems. For our convenience, its evolution can be divided 
into various phases: now we are in the matter-dominated phase, but in 
the past there was certainly a radiation-dominated phase, and an 
explosive phase of very high curvature and density, that we may call 
``big bang". I cannot find any convincing reason, however, to believe that 
before the big bang there was ``nothing". 

To make an analogy with another physical system, consider for instance 
the $\b$-decay of a neutron. The initial neutron is transformed into a 
proton, an electron and a neutrino. For these three particles, the decay 
process is a sort of ``big bang" which marks the beginning of their 
existence. This does not means, however, that these three particles 
spring out of nothing: initially, the system was in a different quantum 
state, representing a neutron. 

In the same way, coming back to 
cosmology, the high curvature big bang phase certainly marks the 
beginning of the present state of the Universe, i.e. of the standard 
Friedmann phase that we can observe today. It seems quite reasonable, 
however, to wonder 
about the state of the Universe preceeding the big bang explosion.
The answer suggested by string theory is illustrated in Fig. 1: our 
present cosmological phase, decelerated, with decreasing curvature, 
might have been complemented by a dual phase, accelerated, with growing 
curvature, which seems natural to call ``pre-big bang". In this 
context, the very early cosmological evolution from the perturbative 
vacuum, i.e. from a configuration with asymptotically flat metric and 
vanishing coupling constant, can be consistently described by the 
lowest-order string effective action, as will be 
discussed in the following Section. 
\begin{figure}[htb]
   \epsfxsize=9cm
   \centerline{\epsfbox{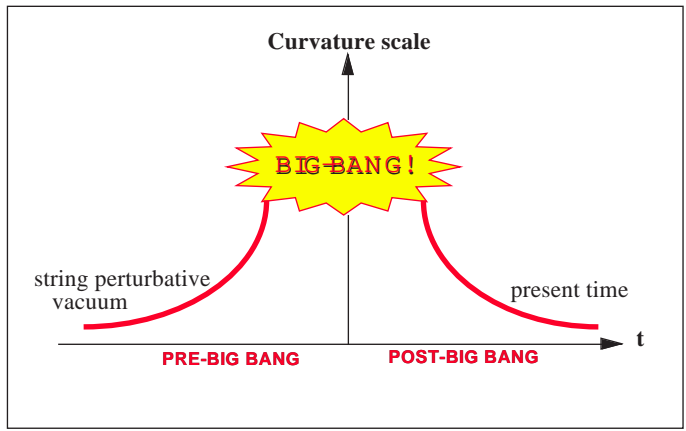}}
   \centerline{\parbox{11.5cm}{\caption{\label{fig:f1}
{\sl Curvature scale versus time in the pre-big bang
scenario. }}}}  
\end{figure}

\vskip 2 cm
\renewcommand{\theequation}{2.\arabic{equation}}
\setcounter{equation}{0}
\section{Low-energy pre-big bang evolution}
\label{sec:2}
\noindent
In the context of the pre-big bang scenario, the global evolution of the 
universe can be conveniently represented in the phase space spanned by
the Hubble factor and by the dilaton kinetic energy. 

Consider in fact the lowest order gravi-dilaton effective action\cite{2},
which in $d$ critical spatial dimensions, and in the string frame, takes the
form
\beq
S = -\frac{1}{2\,\lambda_s^{d-1}}\,\int\,d^{d+1}x\,\sqrt{|g|}\,e^{-\phi}
\,\left(R+\partial_{\mu}\phi\partial^{\mu}\phi \right).\label{21}
\eeq
($\lambda_s\equiv (\ap)^{1/2}$ is the basic string length parameter). For
an isotropic and spatially flat background (but we can also consider more
general initial conditions, spatially curved\cite{2a} and
inhomogeneous\cite{2b}),  the asymptotic solutions approaching the
singularity can be written
as\cite{3}:
\beq
a(t)= (\mp t)^{\mp1/\sqrt d} , ~~~~~~~~~~~~~~
\fb (t)= -\ln (\mp t) , 
\label{22}
\eeq
where $a$ is the scale factor, and $\fb$ is the so-called ``shifted" dilaton:
\beq
\fb=\phi-{d}\ln a-\ln\,\int\,{d^dx\over\lambda_s^d }
\label{23}
\eeq
(we are assuming spatial sections of finite volume). Such solutions are
characterized by four branches, depending on the range of time, and on
the power of the scale factor. As illustrated in Fig. 2, these solutions are
represented by the bisecting lines in the plane spanned by $\sqrt d H$ and
$\dot{\fb}$ ($H=\dot a/a$, and a dot denotes differentiation with respect
to the cosmic time $t$). 

The string perturbative vacuum is characterized by $H=0=\dot{\fb}$, and
corresponds to the origin of that plane. In the limits $|H| \ra \infty$, 
$|\dot{\fb}| \ra \infty$, the background approaches a curvature
singularity. The four branches of the solution describe expansion or
contraction depending on the sign of $H$, and represent a pre-big bang
configuration (evolving towards the curvature singularity) or a post-big
bang configuration (evolving from the singularity towards an
asymptotically flat spacetime), depending on the sign of $\dot{\fb}$. 
Notice that, in the isotropic case, only the branch called ``expanding
pre-big bang" in Fig. 2 describes a true evolution from the perturbative
vacuum (i.e. from a state of zero string coupling), because in the
contracting branch the true dilaton $\phi$ is decreasing\cite{1}, 
and the string coupling $e^\phi$ is also decreasing. 

The four branches of the solution are connected  by $T$-duality
transformations\cite{3}:
\beq
a \ra \ti a = a^{-1} , \,\,\,\,\,\,\,\,\,\,\,\,\,\,\,\,\,\,
\fb \ra \fb , 
\label{24}
\eeq
and $t$-reversal transformations:
\beq
a(t) \ra a(-t) , \,\,\,\,\,\,\,\,\,\,\,\,\,\,\,\,\,\,
\fb(t) \ra \fb (-t) . 
\label{25}
\eeq
A monotonic (expanding or contracting) transition from pre- to post-big
bang thus requires a combination of both $T$ and $t$ transformations, as
illustrated in Fig. 2. Smooth self-dual solutions, characterized by $a(t)=
a^{-1}(-t)$, and defined over the whole time range $-\infty \leq t \leq
+\infty$, would automatically connect the pre- and post-big bang regime,
avoiding the curvature singularity. Such solutions are possible, but only at
the price of adding to the action an ``ad-hoc", 
non-local potential for the shifted dilaton\cite{1,4}. 
\begin{figure}[htb]
   \epsfxsize=9cm
   \centerline{\epsfbox{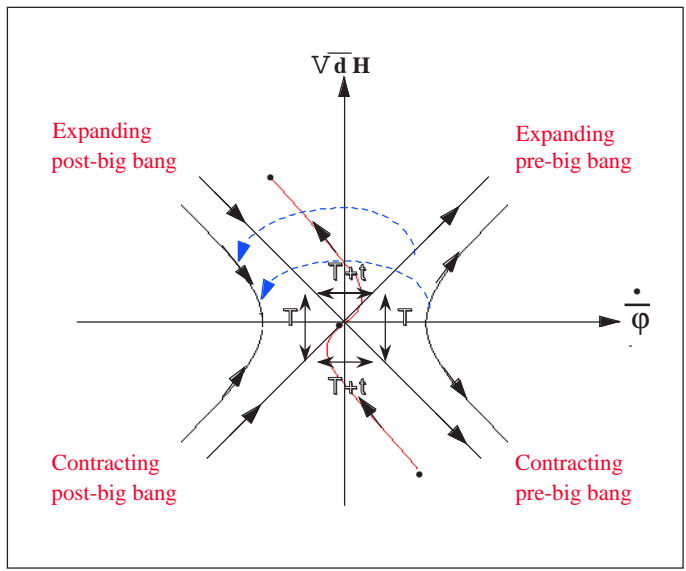}}
   \centerline{\parbox{11.5cm}{\caption{\label{fig:f2}
{\sl Global view of the possible cosmological evolution from and towards
the string perturbative vacuum. }}}}
\end{figure}

The simplest potential is the one induced by a cosmological constant,
$V(\fb)=\La$. In that case the solutions are characterized by the
condition\cite{3}
\beq
{\dot{\fb}}~^2-(\sqrt d H )^2 =\La ,
\label{26}
\eeq
representing an hyperbola in the plane of Fig. 2, and the initial vacuum is
shifted to a state of flat metric and linearly evolving dilaton. The solution
is still characterized by four branches disconnected by a curvature
singularity, so that a transition from the pre- to the post-big bang 
phase remains
classically forbidden. It may be allowed, however, at a quantum level. 

If we compute, with the Wheeler-De Witt (WDW) equation\cite{5}, the
probability of transition to a post-big bang phase with $\La \not= 0$, we
find indeed that such a probability is finite and non-vanishing, in spite of
the presence of a curvature singularity disconnecting, classically, the two
regimes\cite{4,6} (such transitions are represented by the dashed curves
of Fig. 1). 

The transition from a state of positive $\dot{\fb}$ to a state of negative  
$\dot{\fb}$ is allowed even classically, and even in the absence of a
dilaton potential, provided we add to the lowest-order action (\ref{21})
the higher-derivative corrections, arising from an expansion of the string
effective action in powers of the curvature (the so-called $\ap$
expansion\cite{7}). Already to first order in $\ap$, we can find indeed
exact solutions which smoothly interpolate between the pre- and
post-big bang regime\cite{8}, connecting the  perturbative vacuum to a
final state of constant curvature and linearly evolving dilaton, as
illustrated by the solid curve of Fig. 2. 

Fig. 2 illustrates the two possibilities (classical, solid line, and quantum,
dashed lines) of transition from pre- to post-big bang, and thus provides a
schematic summary of my subsequent discussion. 

\renewcommand{\theequation}{3.\arabic{equation}}
\setcounter{equation}{0}
\section{Quantum string cosmology}
\label{sec:3}
\noindent
The quantum approach to the transition is based on the WDW equation,
obtained from the low-energy string effective action. All the 
standard quantum cosmology problems (time parameter, 
probability interpretation, 
semiclassical limit) remain, except perhaps the problem of quantum
ordering, since the ordering is fixed by the duality symmetry\cite{4}. The
aim of this approach is to compare, and possibly contrast, the quantum
results obtained in the context of the standard cosmological scenario, with
the results obtained, with the same method and the same assumptions, in
the context of the pre-big bang scenario. 

The main difference between the two scenarios is qualitatively illustrated
in Fig. 3, in which we compare classical cosmology, quantum cosmology with
tunnelling boundary conditions\cite{9,10}, and quantum string cosmology
for the pre-big bang scenario. While in classical cosmology the initial
singularity is unavoidable, in quantum cosmology the Universe may avoid
the singularity, emerging from the quantum regime through a tunnelling
process. The natural suggestion of string theory is that tunnelling in not
``from nothing"\cite{9}, but from the preceeding pre-big bang phase. Also,  
tunnelling in not through a curvature singularity, but through a string
phase of high (but finite) curvature. 
\begin{figure}[htb]
   \epsfxsize=8cm
   \centerline{\epsfbox{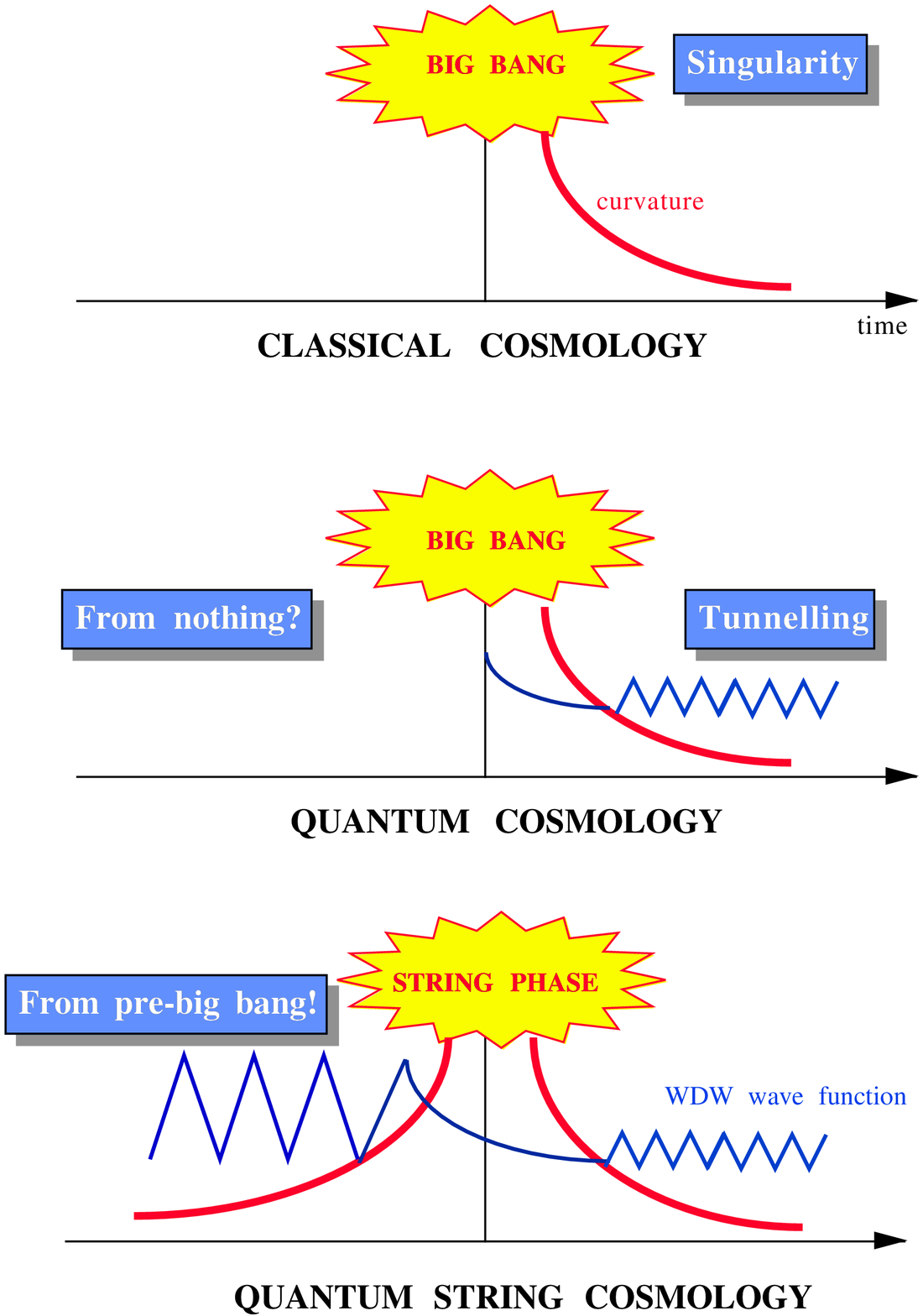}}
\vskip 1cm
   \centerline{\parbox{11.5cm}{\caption{\label{fig:f3}
{\sl A qualitative comparison of classical cosmology, quantum cosmology
with tunnelling boundary conditions, and quantum string cosmology for the
pre-big bang scenario. }}}} 
\end{figure}

\subsection{Quantum transition from pre- to post big bang}
\noindent
For a sort, but self-contained, illustration of how this program
could be implemented, we can start with the simplest gravi-dilaton action
(\ref{21}), supplemented by a dilaton potential $V(\phi)$, in $d=3$ isotropic
dimensions:
\beq
S=\frac{\lambda_s}{2}\,\int\,dt\,{e^{-\fb}\over N}\,
\left(\dot{\beta}^2-\dot{\fb}^2-
N^2\,V \right) . 
\label{31}
\eeq
$N$ is the usual lapse function, and $\b=\sqrt 3 \ln a$.  
The variation of the lapse, in the cosmic time gauge $N=1$, gives the
Hamiltonian constraint
\beq
\Pi^2_{\beta}-\Pi^2_{\fb}
+\lambda_s^2\,V(\b,\fb)\,e^{-2\,\fb}=0~,
\label{32}
\eeq
and the WDW equation
\beq
\left[ \pa_{\fb}^2-\pa_\b^2 +\lambda_s^2\,V(\b,\fb)\,e^{-2\,\fb}\right]
\psi =0,
\label{33}
\eeq
where we have introduced the canonical momenta:
\beq
\Pi_{\beta}={\da S\over \da\dot{\beta}}=
\lambda_s\,\dot{\beta}\,e^{-\fb} , ~~~~~~~~~~~~
\Pi_{\fb}={\da S\over \da\dot{\fb}}=
-\lambda_s\,\dot{\fb}\,e^{-\fb} .
\label{34}
\eeq

In the absence of a dilaton potential the WDW equation 
(\ref{33}) reduces to the
massless Klein-Gordon equation, and we recover a plane wave
representation of the four branches of the classical solution, 
$\psi_{\b}^{(\pm)}\psi_{\fb}^{(\pm)} \sim e^{\pm ik \b \pm ik \fb}$, where
\beq
\Pi_{\b}\,\psi_{\b}^{(\pm)}= \pm k \,\psi_{\b}^{(\pm)} , ~~~~~~~~~~~
\Pi_{\fb}\,\psi_{\fb}^{(\pm)}= \pm k \,\psi_{\fb}^{(\pm)}.
\label{35}
\eeq
They corresponds to expansion ($\Pi_\b >0$), contraction ($\Pi_\b <0$),
growing dilaton ($\Pi_{\fb}<0$), decreasing dilaton ($\Pi_{\fb} >0$),
according to eq. (\ref{34}). A transition from a state of pre-big bang
expansion to a state of post-big bang expansion thus corresponds, in this
representation, to a transition from a state of positive momentum 
$\Pi_\b$
and negative momentum $\Pi_{\fb}$, to a state with positive $\Pi_{\b}$ 
and $\Pi_{\fb}$. In other words, the transition corresponds to a monotonic
evolution along the $\b$ direction, and to a reflection along the $\fb$
direction. 
\begin{figure}[htb]
   \epsfxsize=8cm
   \centerline{\epsfbox{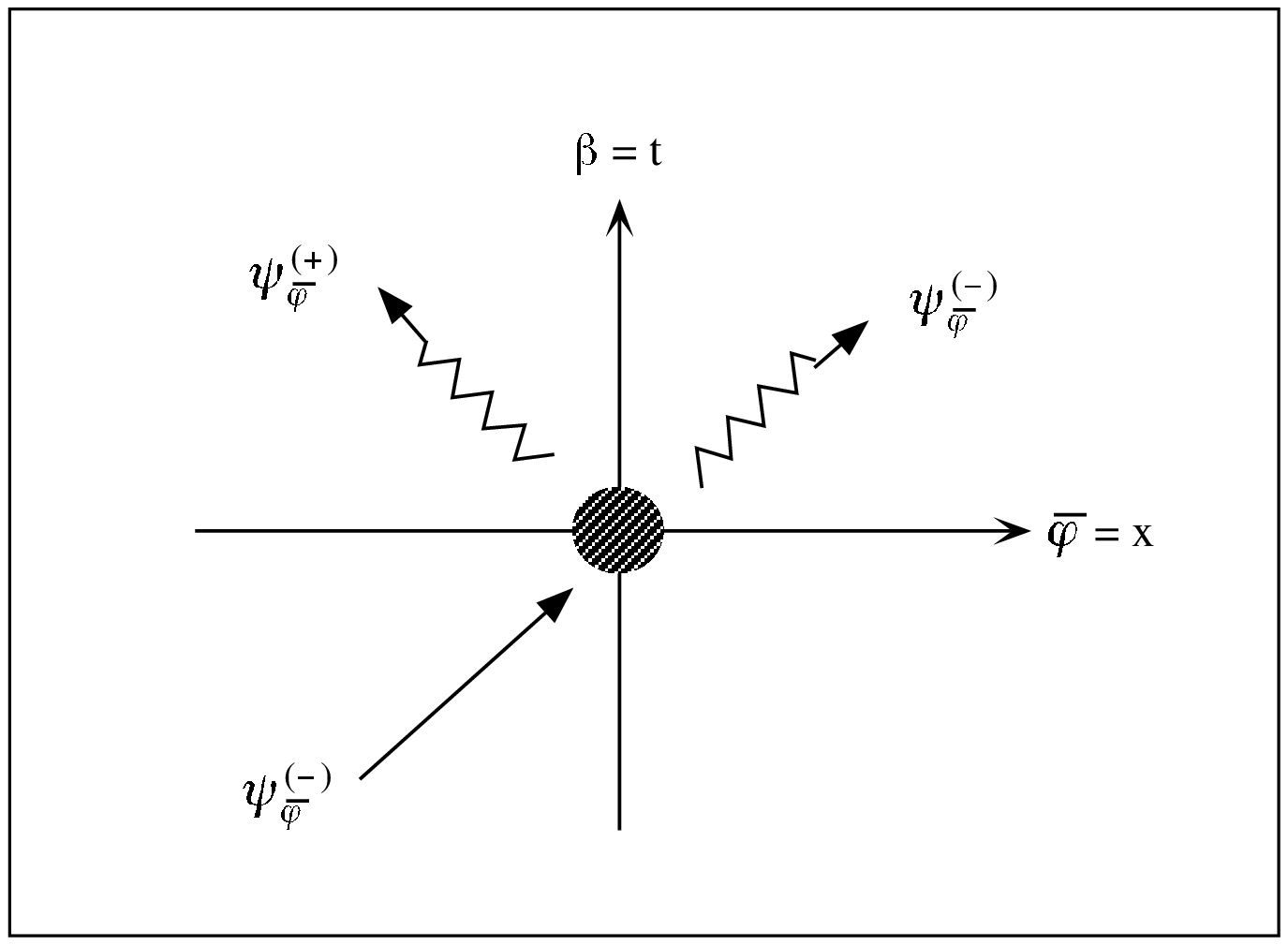}}
   \centerline{\parbox{11.5cm}{\caption{\label{fig:f4}
{\sl Transition from pre- to post-big bang represented as a spacelike
reflection of the wave function, in the minisuperspace spanned by $\b$
and $\fb$. }}}}  
\end{figure}

This suggests to look at the transition as at a scattering process\cite{4,6}
in the minisuperspace spanned by $\b$ and $\fb$, in which $\b$ plays the
role of the time-like coordinate, and $\fb$ plays the
role of the space-like coordinate, as illustrated in Fig. 4. The initial pre-big
bang state, $\psi \sim  e^{+ ik \fb - ik \b}$, is partially transmitted towards
the singularity, and partially reflected back to a post-big bang
configuration, $\psi \sim  e^{- ik \fb - ik \b}$. The transition probability is
controlled by the reflection coefficient, $R=|\psi_{\fb}^{(+)}|^2/
|\psi_{\fb}^{(-)}|^2$ (see Ref. [13] for a 
rigorous definition of scalar products in the appropriate Hilbert space). 

The boundary conditions, in this context, are unambiguously fixed by the
choice of the perturbative vacuum as the initial state of the Universe. It
may be noted that, with this choice, there are only outgoing waves at the
singular spacetime boundary ($\fb \ra +\infty$), just like in the case of
tunnelling boundary conditions\cite{9}. We may thus look at the process as
at a sort of ``tunnelling from the string perturbative vacuum", even if, in
this minisuperspace representation, the WDW wave function is actually
reflected. 

When the reflection is induced by a cosmological constant, $V=\La$, 
we have checked\cite{6} 
that the transition probability is indeed very similar to the
tunnelling probability. Both are exponentially suppressed, unless the
cosmological constant is very large and the proper size of the transition
volume is very small in string units. The only basic difference is that, in a
string cosmology context, the coupling constant is controlled by the
dilaton and it is thus running in Planckian units. As a consequence, the
Universe tends to emerge from the pre-big bang phase in the strong
coupling regime, since 
the transition probability has a typical instanton-like
behaviour $\sim \exp (-g_s^{-2})$, where $g_s= \exp (\phi/2)$ is the string
coupling parameter.

It is important to stress, finally, that the WDW eq. (\ref{33}) is free from
problems of operator ordering. Indeed, thanks to the duality symmetry of
the effective action, the WDW Hamiltonian is associated to a globally flat
minisuperspace. 

For a more general discussion of this point, we may add to the effective
action (\ref{31}) a non-trivial antisymmetric tensor (also called torsion)
background, $B_{\mu\nu}\not= 0$. The kinetic part of the action may then
be written in compact form as
\beq
S=-{\ls\over 2}\int dt e^{-\fb}\left[(\dot{\fb})^2+{1\over 8}{\rm Tr}
~\dot M(M^{-1})\dot{}\right] , 
\label{36}
\eeq
where $M$ is a symmetric $2d \times 2d$ matrix, including the spatial
part of the metric, $G\equiv g_{ij}$, and of the torsion, $B\equiv B_{ij}$:
\beq
M=\pmatrix{G^{-1} & -G^{-1}B \cr
BG^{-1} & G-BG^{-1}B \cr} .
\label{37}
\eeq
This action is invariant under global $O(d,d)$ transformations\cite{11}
which leave invariant the shifted dilaton, 
\beq
\fb \ra \fb , ~~~~~~~~~~ M\ra \Om^T M \Om , 
\label{38}
\eeq
where
\beq
\Om^T\eta \Om =\eta, ~~~~~~~~~~ \eta =
\pmatrix{0 & I \cr I & 0 \cr} .
\label{39}
\eeq

The Hamiltonian associated to  torsion-graviton background,
\beq
H \sim {\rm Tr}~(M~\Pi_M M~\Pi_M), ~~~~~~~~~\Pi_M = \da S/\da M, 
\label{310}
\eeq
would seem to have ordering problems, because $[M, \Pi_M] \not= 0$.
However, thanks to the $O(d,d)$ properties of $M$, 
\beq
M\eta M =\eta , 
\label{311}
\eeq
we can always rewrite the kinetic part of the action in terms of the flat
$O(d,d)$ metric $\eta$:
\beq
{\rm Tr}~ \dot M(M^{-1}) \dot{}=
{\rm Tr}~ (\dot M\eta)^2.
\label {312}
\eeq
The corresponding Hamiltonian
\beq
H\sim {\rm Tr}~(\eta~\Pi_M~ \eta~\Pi_M)
\label{313}
\eeq
has a flat metric in momentum space, with no ordering problems for the
corresponding WDW equation. 

It may be noted that, for a general curvilinear parametrization of
superspace, the ordering fixed by the duality symmetry is equivalent to
the ordering fixed by the requirement of reparametrization
invariance\cite{4}. It is crucial, for this equivalence, the fact that there
are no contributions to the ordered Hamiltonian from the scalar curvature
of minisuperspace, because minisuperspace is globally flat. 

\subsection{Quantum transition from expansion to contraction}
\noindent
The transition discussed in the previous Section was characterized by a
monotonic evolution of the scale factor. In the context of the pre-big bang
scenario, however, it is also important to consider the possibility of
transitions from expansion to contraction. Indeed, if we start with a
higher-dimensional and anisotropic perturbative vacuum, the initial
growth of the dilaton requires a large enough number of expanding
dimensions\cite{1}, in general larger than $3$. 
On the other hand, a model of
dynamical dimensional reduction requires only three expanding dimensions,
with all the other dimensions contracting down to a final compactification
scale. The late-time transition of an appropriate 
subsection of the  spatial manifold, from the
expanding pre-big bang phase to the contracting pre-big bang phase, may
thus be useful to implement a realistic cosmological scenario. 

If the dilaton potential depends only on $\fb$, the action (\ref{31}) is
duality-invariant, the momentum along $\b$ is conserved, $[\Pi_\b, H]=0$,
the evolution of the scale factor is monotonic, and the transition from
expansion to contraction is classically forbidden. Such a transition is
allowed at the quantum level, however, where it can be represented (in a
third quantization formalism) as a process of pair production (pairs of
universes, in this case), out of the string perturbative vacuum\cite{12}.
This process is described by the same action, same Hamiltonian, same
minisuperspace as before, but with different boundary conditions, and
with a $90$ degree rotation of the time axis (see Ref. [16] for a
classification of the different types of scattering of the perturbative
vacuum in the ($\b,\fb$) minisuperspace). 
\begin{figure}[htb]
   \epsfxsize=8cm
   \centerline{\epsfbox{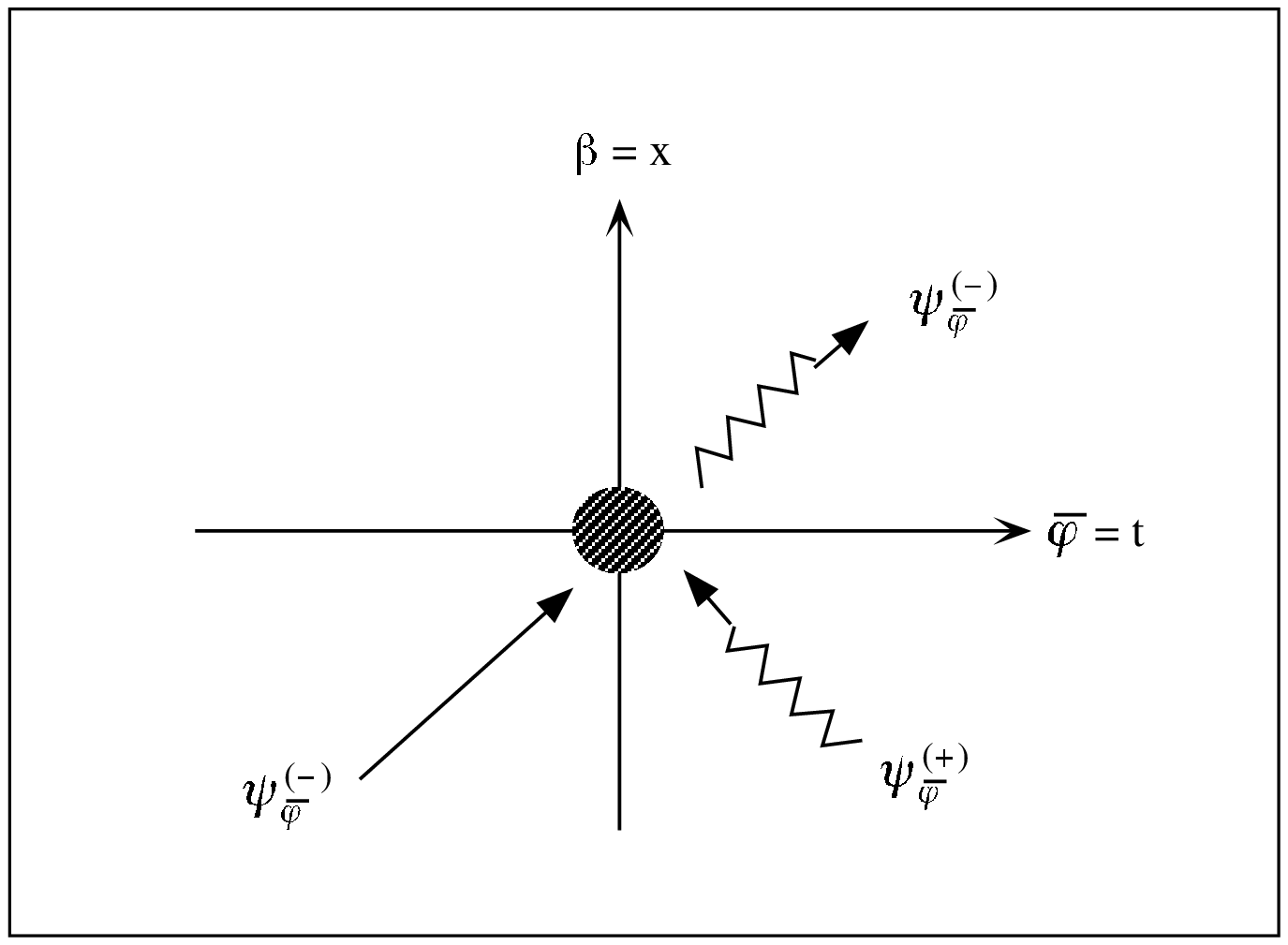}}
   \centerline{\parbox{11.5cm}{\caption{\label{fig:f5}
{\sl Production from the vacuum of a universe--anti-universe pair, one
expanding, the other contracting. }}}}   
\end{figure}

Consider in fact the configuration illustrated in Fig. 5, where we have one
incoming wave in the weak coupling regime $\fb \ra -\infty$, and two
waves, one incoming, the other outgoing, in the strong coupling regime 
$\fb \ra +\infty$. All the asymptotic states appearing in Fig. 5 are
eigenstates of $\Pi_\b$, with positive momentum $k>0$. However, if $\fb$ is
the time-like coordinate, the final state is a mixture of positive and
negative frequency modes, $\psi_{\fb}^{(\pm)}$. In a quantum field theory
context, this configuration describes a Bogoliubov process of pair creation
out of the vacuum, in which the negative frequency mode is re-interpreted
as an anti-particle of positive frequency (i.e. positive energy) and opposite
spatial momentum. 

In our case the spatial momentum is $\Pi_\b$, and  a negative value of
$\Pi_\b$ corresponds to a contraction. The final state of the process
illustrated in Fig. 5 thus describes, in a third quantization formalism, the
production from the vacuum of a universe--anti-universe pair, one
expanding, the other contracting. 

With such a boundary conditions, the process cannot be interpreted as a
tunnelling. It can be interpreted, instead, as an ``anti-tunnelling from the
string perturbative vacuum", i.e  as a process in which the WDW wave
function is parametrically amplified in superspace\cite{12}. The probability
of the process is indeed controlled by the Bogoliubov coefficient which
weights the anti-universe content of the final state, and which 
becomes, in the parametric amplification regime,  
the inverse of the quantum mechanical transmission coefficient
associated to the scattering process. 

\vskip 2 cm
\renewcommand{\theequation}{4.\arabic{equation}}
\setcounter{equation}{0}
\section{Classical evolution from pre- to post-big bang}
\label{sec:4}
\noindent
A classical description of the transition, in terms of an exact solution 
connecting in a smooth way the pre- and 
post-big bang regime, is known to be excluded in the context of the 
lowest-order string effective action, for any local (and realistic) 
dilaton potential\cite{14}. It may be allowed, however, when higher 
order corrections to the action are taken into account. 

In string theory there are two types of higher order corrections: 
higher-derivative terms, appearing in the so-called 
$\ap$-expansion\cite{7}, and and higher loops in the string coupling 
parameter $g_s= e^{\phi/2}$. The first is a truly ``stringy" effect, 
due to the presence of the minimal, fundamental length 
$\la_s=(\ap)^{1/2}$; the second is a quantum effect, controlled by the 
expectation value of the dilaton field. 

Both types of corrections are probably required for a complete and 
successful description of the transition. For pedagogical reasons, 
however, it is better to discuss their effects separately, starting with 
the $\ap$ corrections that are probably the first to come into play when 
the background evolves from the perturbative vacuum\cite{8}. 

\subsection{Higher-derivative corrections}
\noindent
To first order in $\ap$, i.e. including all terms required by the 
gravi-dilaton sector of string theory up to four derivatives in the 
tree-level action, the action can be written as\cite{7}:
\beq
S=-{1\over 2\la_s^{d-1}}\int d^{d+1}x \sqrt{|g|}e^{-\phi} \left[ R+
(\nabla \phi)^2-{\ap \over 4} \left(R^2_{GB} - 
(\nabla \phi)^4\right)\right] ,
\label{41}
\eeq
where $R^2_{GB} 
\equiv R_{\mu\nu\a\b}^2-4  R_{\mu\nu}^2+R^2$  
is the usual Gauss-Bonnet invariant. Note that there are no free 
parameters, except a possible number of order unity in front of the 
$\ap$ corrections, depending on the particular (super)string theory 
(bosonic, heterotic, ...) adopted. We have chosen here a convenient 
field-redefinition that eliminates higher-than-second derivatives from 
the equations of motion, but at the price of introducing 
dilaton-dependent $\ap$ corrections. 

\begin{figure}[htb]
   \epsfxsize=6cm
   \centerline{\epsfbox{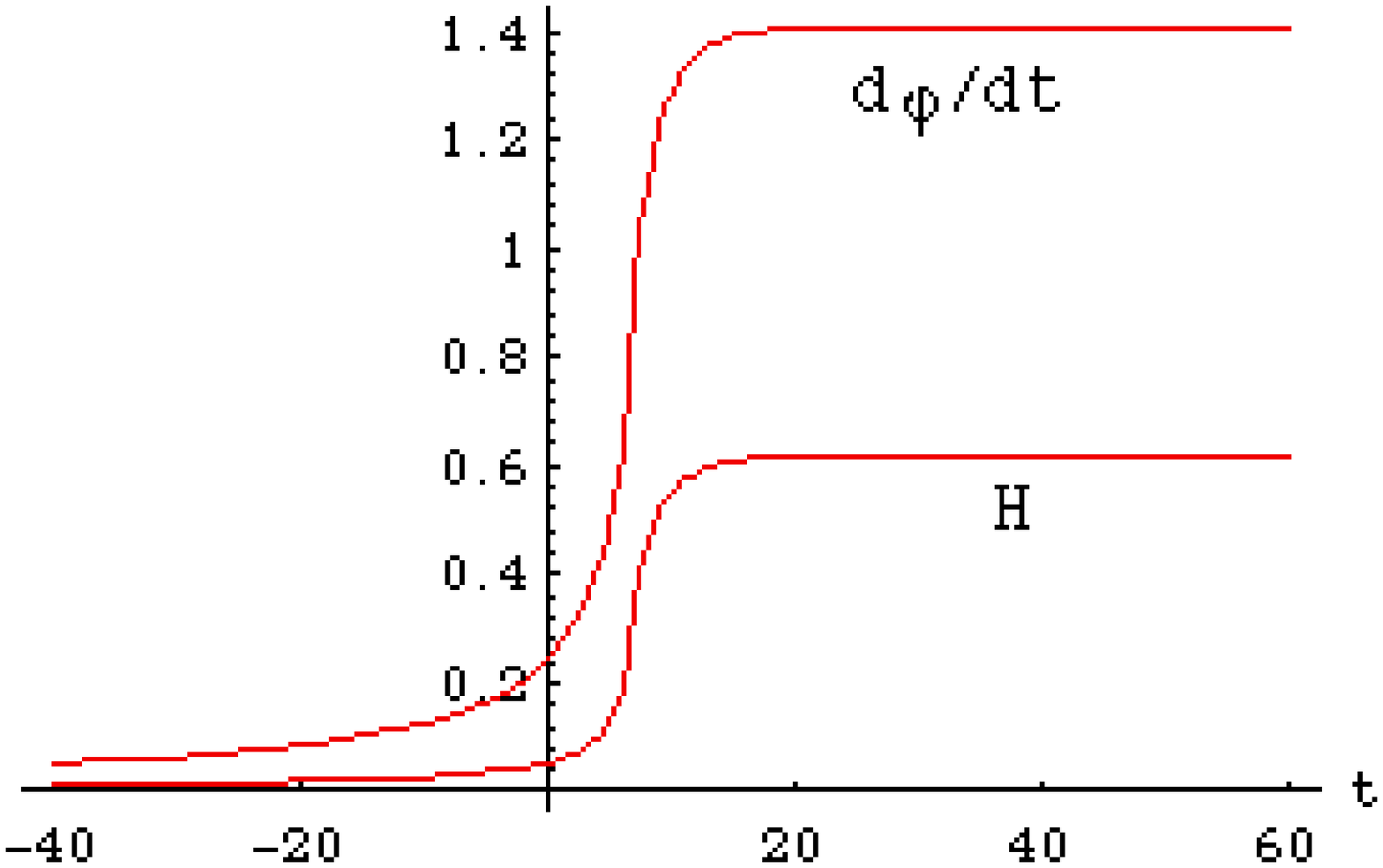}}
   \centerline{\parbox{11.5cm}{\caption{\label{fig:f6}
{\sl Numerical integration of the equations of motion for the action 
(\ref{41}), in $d=3$, and with pre-big bang initial conditions.}}}} 
\end{figure}

A numerical integration of the equations of motion (in any number of 
dimensions) shows that, already to this order in $\ap$, the effect of 
the higher-derivative terms is to contrast the growth of the curvature 
and of the dilaton, leading the background to a final regime in which 
the curvature is constant, and the dilaton is linearly growing (in 
cosmic time). This is clearly illustrated in Fig. 6, which shows the 
result of a numerical integration of the equations of motion, with the 
perturbative vacuum ($H=0=\dot\phi$) as initial conditions at $t\ra 
-\infty$. The plot refers to the case $d=3$, but the result is 
qualitatively the same\cite{8} for any $d$. 

It is important to stress that such a 
state of frozen curvature and linear dilaton may represent a solution to 
all orders in $\ap$ of the tree-level action. Indeed, the complete set 
of the string $\sigma$-model $\b$-function equations, 
for a Bianchi-type I gravi-dilaton background, has been shown to reduce, 
to all orders, 
to an algebraic set of $d+1$ equations in $d+1$ unknowns, representing 
the (anisotropic) linear rate of growth of the log of the metric and of 
the string coupling\cite{8}. The existence of real solutions for this 
system is a sufficient condition for the existence of an exact (in the 
sense of conformal field theory, i.e. to all orders in $\ap$) solution 
with constant curvature and linearly evolving dilaton. 

It may be noted that a solution with $H=$ const, $\dot \phi=$ const, 
is in general allowed in many higher-derivative models of gravity. Such 
a fixed point of the cosmological equations, however, is in general 
disconnected from the trivial fixed point $H=0=\dot \phi$ (the 
perturbative vacuum, in this case) by a singularity, or by an unphysical 
region in which the curvature becomes imaginary. For the action 
(\ref{41}), on 
the contrary, the constant fixed point is a late time time attractor for 
all isotropic backgrounds emerging from the string perturbative 
vacuum\cite{8}. 

This attraction property, unfortunately, is not invariant under field 
redefinition, as we have checked, until the action is truncated at a 
given finite order in $\ap$. Also, and most important, the growth of the
curvature is stopped, but the transition is not completely performed.
Indeed, the final fixed point is in the post-big bang regime
$\dot{\fb}<0$, as illustrated
in Fig. 2, but the transition cannot proceed further towards the lower
curvature regime, without including additional corrections or an
appropriate dilaton potential in the effective action. 
\begin{figure}[htb]
   \epsfxsize=8cm
   \centerline{\epsfbox{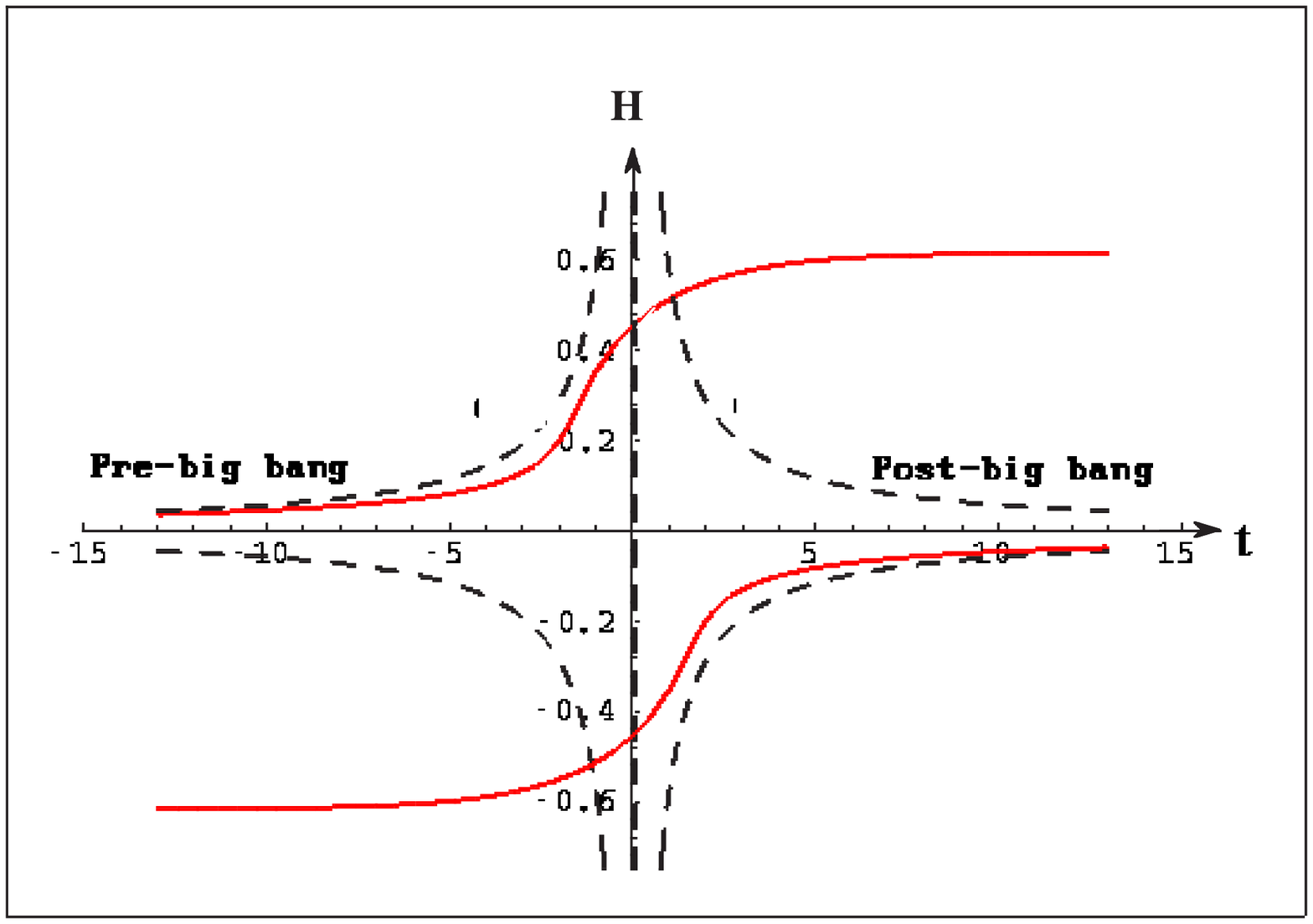}}
   \centerline{\parbox{11.5cm}{\caption{\label{fig:f7}
{\sl Time-reversal symmetry of the regularized solution (the dashed
curves represent the four branches of the lowest order, singular
solution).}}}}   
\end{figure}

The reason for this incompleteness is that the action (\ref{41}) is invariant
under time reflections, but is not duality-invariant. As a consequence,
what is regularized is the expanding pre-big bang branch
of the lowest order solution, and its time-symmetric counterpart, the 
{\it contracting post-big bang} branch. This is illustrated 
in Fig. 7,  
where we plot the evolution in time of the curvature scale, according to
a numerical integration of the equations of motion. The dashed curves
are the four branches of the corresponding lowest-order, singular
solution. The expanding post-big bang branch remains singular, and
cannot be smoothly connected to the regularized pre-big bang branch.
More corrections are to be added to the effective action.

\subsection{Loop corrections}
\noindent
One-loop corrections to the string cosmology equations have been
discussed by various authors, in $D=2$ and $D=4$ spacetime dimensions.
In two dimensions\cite{16}, the one-loop corrections to the so-called
CGHS model\cite{17} lead to the action
\beq
S= S_{{\rm tree}}+ 
{k\over 2}\int d^2x \sqrt{-g}\left(R\Box^{-2}R+\ep\phi R\right), 
\label{42}
\eeq
where $S_{{\rm tree}}$ is the tree-level action, and $k$ is a
dimensionless parameter that depends on the number of conformal
scalar field present in the model. Notice that the trace-anomaly term 
$R\Box^{-2}R$ has been supplemented by the local covariant
counterterm $\ep\phi R$, that one is free to add to preserve classical
symmetries\cite{17a}. 

In four dimensions, the dimensionally reduced one-loop action can be
written as\cite{15}
\beq
S= S_{{\rm tree}}- \ap \da \int d^4 x \sqrt{-g} \xi (\sg) F\left[R^2, 
(\pa_\mu \phi)^4, R  (\pa_\mu \phi)^2, ....\right]
\label{43}
\eeq
where $\da$ is a model-dependent number of the order of unity, $\xi
(\sg)\approx - (2 \pi/3) \cosh \sg$ is a function of the modulus field
$\sg$ parametrizing the size of the internal compactified manifold, and
$F$ is a complicated four-derivative function of the metric and of the
dilaton, that in the Einstein frame factorizes into the Gauss-Bonnet 
form 
times a dilaton-dependent function. The tree-level action, in eq. 
(\ref{43}), includes
the modulus field and the first order $\ap$ corrections. 

Both the analytical and numerical solutions of the actions (\ref{42})
and (\ref{43}) show that the effect of loop corrections is to induce a
bounce in the curvature\cite{16,15}: the curvature grows, reaches a
maximum and then decreases, with a typical ``bell-like" shape which
mimics the effects of a transition from pre- to post-big bang. 

The dilaton, however, remains growing from $-\infty$ to $+\infty$. It is
true that a monotonic evolution of the dilaton from weak to strong
coupling is equivalent, via S-duality transformations, to a smooth
interpolation between two different weak coupling regimes. The final
value of the string coupling should go to a finite constant, however,
and not to zero. To this aim the dilaton has to be stopped, and this
probably requires the addition of a realistic, non-perturbative
potential. 

The problem with the cosmological solutions of the one-loop corrected
action, at least for the cases that we have analyzed, is that the final
background configuration is still characterized by $\dot{\fb}>0$. This
means, in other words, that there is no branch changing: the
background is still in the pre-big bang regime (in spite of the fact that
the curvature is decreasing), and the dilaton is not ready to be
attracted to any stable minimum of the potential\cite{14}.

We may thus conclude, to the best of our present understanding of
string cosmology, that $\ap$ and loop corrections can separately
implement different aspects of the transition from pre- to post-big
bang. They are both required, however, together with an appropriate
non-perturbative potential, for the description of a complete
transition. 

\vskip 1 cm

\renewcommand{\theequation}{5.\arabic{equation}}
\setcounter{equation}{0}
\section{Conclusion}
\label{sec:5}
\noindent
String theory can provide (at least in principle) a consistent description
of the Universe at all curvature scale. The duality symmetries of string
theory, in particular, suggest a cosmological scenario in which the
Universe evolves from the string perturbative vacuum, through a
pre-big bang phase characterized by an accelerated growth of the
curvature and of the string coupling (controlled by the dilaton). 

I do not conceal that what I have presented here is no more than an
approximate sketch of a complete and possibly realistic cosmological
scenario. A lot of work is still needed to clarify all the aspects of this
scenario, and in particular the dynamical details of the transition to the
standard cosmological regime. 

It seems to me, however, that such a work is worth to be done,
because the study of this pre-big bang scenario, and of the associated 
non-standard phenomenology\cite{1,18}, may provide an efficient
way to test string theory as well as alternative
models of Planck scale physics.

\vspace{1cm}
{\it Acknowledgments:\/} I am grateful to  A. Buonanno, M. Maggiore,
J. Maharana, R. Ricci, C. Ungarelli and G. Veneziano for many useful
discussions, and for a fruitful collaboration. I wish to thank also H. J.
De Vega and N. Sanchez for their kind  invitation, and for the perfect
organization of this Euroconference.

\end{document}